\documentclass[preprint,12pt]{elsarticle}
\usepackage{graphics}

\usepackage{amssymb}

\journal{Nuclear Physics A}

\begin{document}

\begin{frontmatter}


\title{Precise half-life measurement of the 10 h isomer in $^{154}$Tb}

\author{Gy.~Gy\"urky\corref{cor1}}
\ead{gyurky@atomki.hu}
\cortext[cor1]{corresponding author}
\author{G.~Rastrepina, Z.~Elekes, J.~Farkas, Zs.~F\"ul\"op, G.G.~Kiss, E.~Somorjai, T.~Sz\H ucs}
\address{Institute of Nuclear Research (ATOMKI), H-4001 Debrecen, POB.51., Hungary}

\begin{abstract}
The precise knowledge of the half-life of the reaction product is of crucial importance for a nuclear reaction cross section measurement carried out with the activation technique. The cross section of the $^{151}$Eu($\alpha$,n)$^{154}$Tb reaction has been measured recently using the activation method, however, the half-life of the 10 h isomer in $^{154}$Tb has a relatively high uncertainty and ambiguous values can be found in the literature. Therefore, the precise half-life of the isomeric state has been measured and found to be T$_{1/2}$\,=\,9.994\,h\,$\pm$\,0.039\,h. With careful analysis of the systematic errors, the uncertainty of this half-life value has been significantly reduced.
\end{abstract}


\end{frontmatter}


\section{Introduction}
\label{sec:introduction}

Modelling of the nucleosynthesis of the heavy elements (e.g. in the astrophysical r- and p-processes) requires knowledge of the rates of thousands of reactions. In lack of experimental data, the reaction rates are usually obtained from statistical model calculations. Statistical models utilize different input parameters which strongly influence the calculated cross sections and hence the reaction rates. Besides the astrophysical conditions, inaccurate nuclear input parameters used in the reaction rate calculations can be responsible for the fact that the nucleosynthesis models are not able to reproduce well the heavy element abundances observed in nature \cite{rap06,rau06}. Therefore, the measurement of the relevant reaction cross sections is of high importance to check the results of statistical model calculations and to help choose the best input parameters.

It has been found that the results of statistical model calculations strongly deviate from the measured cross section values in the case of reactions involving alpha particles. The reason of this discrepancy is that the calculated cross sections show a strong dependence on the alpha-nucleus optical potential and this potential is ambiguous at the astrophysically relevant low energies (see e.g. \cite{gyu06}). Therefore, the need for experimental cross sections is more pronounced in the case of alpha-induced reactions. 

Realizing this need, the cross sections of the $^{151}$Eu($\alpha,\gamma$)$^{155}$Tb and $^{151}$Eu($\alpha$,n)$^{154}$Tb reactions have been measured at astrophysically relevant energies using the activation technique. In this method, the cross section is deduced from the off-line measurement of the reaction products,  $^{155}$Tb and $^{154}$Tb. The half-life of the reaction products enter into the calculation of the cross section, and thus the uncertainty of the half-life values directly contributes to the uncertainty of the obtained cross sections.

$^{155}$Tb does not have any long lived isomer and the half-life of its ground state is relatively well known: 5.32\,$\pm$\,0.06\,d \cite{NDS104}. The situation for $^{154}$Tb is, however, more complicated \cite{NDS85}. Its ground state decays by $\beta^+$ emission and electron capture to $^{154}$Gd with a half-life of T$_{1/2}^{\rm g.s.}$\,=\,21.5\,$\pm$\,0.4\,h. $^{154}$Tb is has two long-lived isomeric states with unknown excitation energies but well established level ordering. The m1 isomer decays both by internal transition to the ground state and by $\beta^+$ and electron capture to $^{154}$Gd with a half-life of T$_{1/2}^{\rm m1}$\,=\,9.4\,$\pm$\,0.4\,h. The m2 isomer decays almost exclusively by $\beta^+$ and electron capture to $^{154}$Gd; there is only a weak internal transition to the m1 state. One can note that the half-life of the m1 isomer has an unusually large uncertainty of more than 4\,\%. This is because ambiguous data can be found in the literature. The measurements can be grouped into two categories. In the first group three half-life values of about 10\,h can be found by Vylov et al. (9.9\,$\pm$\,0.1\,h \cite{Vyl72}), by Berkes et al. (9.8\,$\pm$\,0.3\,h \cite{Ber83}) and by Nedovesov et al. (9.9\,$\pm$\,0.4\,h \cite{Ned76}). In the second group, two less precise and lower half-life values can be found by Lau and Hogan (9.0\,$\pm$\,0.5\,h \cite{Lau73}) and Sousa et al. (9.0\,$\pm$\,1.0\,h \cite{Sou75}). In the compilation \cite{NDS85}, a recommended value of 9.4\,$\pm$\,0.4\,h is given with a remark: {\it ''An unweighted average was chosen because the two more precise values are also the highest and may include an unaccounted for contribution from the 22.7 {\rm h} isomer``.} If one accepts that some of the available half-life measurements involve a hidden systematic error, the half-life of the m1 isomer can have an uncertainty of up to 10\%. Since the preliminary analysis of the $^{151}$Eu($\alpha$,n)$^{154}$Tb cross section measurement shows that the m1 isomer is by far the most strongly populated state in $^{154}$Tb, the uncertainty of the m1 isomer half-life substantially influences the obtained cross section of the $^{151}$Eu($\alpha$,n)$^{154}$Tb reaction. Therefore, a high precision experiment has been carried out aiming at the determination of the m1 isomer half-life. In Section \ref{sec:experimental} the details of the experiment are presented. Section \ref{sec:analysis} shows the data analysis and the results are presented in Section \ref{sec:results}.

\section{Experimental technique}
\label{sec:experimental}
\subsection{Source preparation}
\label{sec:sourceprep}

$^{154}$Tb sources have been produced by the $^{151}$Eu($\alpha$,n)$^{154}$Tb reaction. The targets have been prepared by evaporating Eu$_2$O$_3$ enriched to 99.2\,\% in $^{151}$Eu onto thin Al foils. The use of enriched $^{151}$Eu target was necessary because the ($\alpha$,n) reaction on the heavier Eu isotope leads to $^{156}$Tb (T$_{1/2}$\,=\,5.35\,$\pm$\,0.1\,d) and the decay of this isotope is followed by a strong $\gamma$-radiation with similar energy to the one from $^{154}$Tb used in the analysis (see Sec. \ref{sec:gammacounting}). A thin ($\approx$ 10\,$\mu$g/cm$^2$) Al protective layer has been evaporated onto each target in order to prevent the produced $^{154}$Tb nuclei from diffusing out of the source during the half-life measurement.

\begin{table}
\centering
\caption{\label{tab:sources} Details of the source preparation}
\begin{tabular}{lcccc}
\hline
Source  & E$_\alpha$ [MeV] & $^{151}$Eu target  & Accumulated\\
 & & thickness & charge\\
 & & [10$^{15}$ atom/cm$^2$] & [$\mu$C]\\
\hline
\#1 & 13.5 & 590 & 8.9$\cdot$10$^4$ \\
\#2 & 14.5 & 580 & 9.1$\cdot$10$^4$ \\
\#3 & 15 & 480 & 8.7$\cdot$10$^4$ \\
\#4 & 15.5 & 260 & 4.2$\cdot$10$^4$ \\
\#5 & 17 & 500 & 2.2$\cdot$10$^4$ \\
\hline
\end{tabular}
\end{table}

The targets have been bombarded by alpha beams from the cyclotron of ATOMKI. The typical beam intensity was 2\,$\mu$A. Five sources have been prepared with different alpha bombarding energies of 13.5, 14.5, 15, 15.5 and 17\,MeV. The irradiation times varied between 5 and 12 hours. In Table \ref{tab:sources}, information about the source preparation can be found.

\subsection{Gamma counting}
\label{sec:gammacounting}
 
The half-life of $^{154}$Tb has been measured by detecting a characteristic $\gamma$-radiation following the decay for an extended period of time. A 40\,\% relative efficiency Canberra n-type HPGe detector has been used for the $\gamma$-counting. In order to reduce the laboratory background, the detector was surrounded by a 5\,cm thick lead shield. The signals from the detector preamplifier have been shaped and amplified by an ORTEC model 671 spectroscopic amplifier. The signals were then fed into an ORTEC model ASPEC-927 multichannel analyzer and the spectra were collected using an ORTEC A65-B32 MAESTRO software. This data acquisition system has a built-in dead time correction, but in order to check the dead time which is crucial for a precise half-life experiment, a pulse generator has been included in the system. For details of the dead time correction see Sec. \ref{sec:analysis}. 

The spectra were stored in every hour and the total length of the counting varied between 21 and 62 hours (more than 6 half-lives) depending on the source activity. Table \ref{tab:counting} lists some details of the $\gamma$-counting.

The decay of the ground state and the two isomeric states of $^{154}$Tb involves the emission of numerous different energy $\gamma$-radiations. According to the work of e.g. Sousa et al. \cite{Sou75}, some of the $\gamma$-transitions can be assigned exclusively to a given state of $^{154}$Tb. In the case of the m1 isomer, the strongest $\gamma$-line which is associated only to this state is at 540.2\,keV. This line has been used for the analysis. All other $\gamma$-lines solely from this isomer are at least one order of magnitude weaker and therefore their inclusion in the analysis did not improve the precision obtained on the 540.2\,keV line.

\begin{table}
\caption{\label{tab:counting} Details of the $\gamma$-counting.}
\begin{minipage}{\textwidth}
\centering
\begin{tabular}{lccc}
\hline
Source  & Source & length of & Initial   \\
 & activity\footnote{The activity of the m1 isomer of $^{154}$Tb at the end of irradiation} & counting & dead time  \\
 & [kBq] & [h] & [\%] \\
\hline
\#1 & 1.4 & 25 & 1.2 \\
\#2 & 4.7 & 22 & 2.8 \\
\#3 & 15 & 21 &  5.2 \\
\#4 & 12 & 53 &  2.6 \\
\#5 & 115 & 62 &  10.1\\
\hline
\end{tabular}
\end{minipage}
\end{table}
 
\begin{figure}
\centering
\resizebox{0.8\textwidth}{!}{\rotatebox{270}{\includegraphics{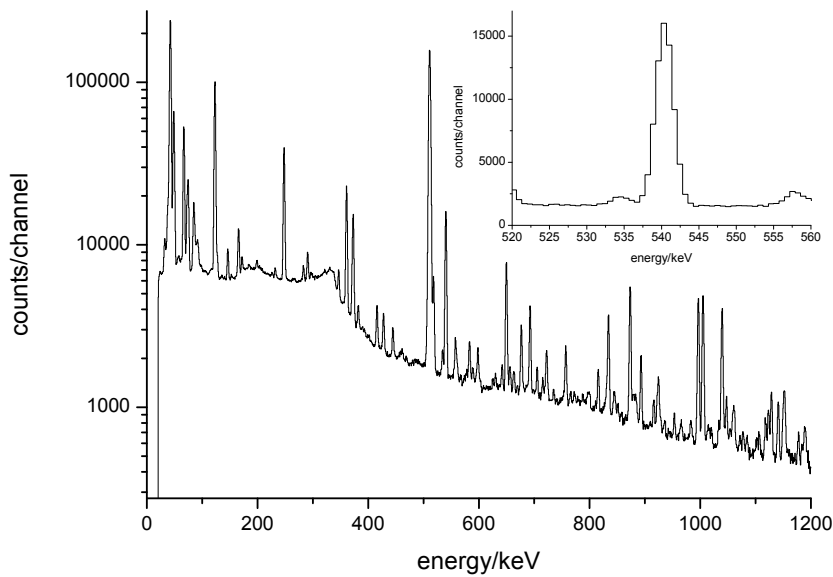}}}
\caption{\label{fig:spectrum} $\gamma$-spectrum measured on source \#2 for one hour. The inset shows the region of the 540.2\,keV peak.}
\end{figure}

Figure \ref{fig:spectrum} shows a typical $\gamma$-spectrum measured on source \#2 roughly in the middle of the counting period for one hour. The very complex decay spectrum of $^{154}$Tb is apparent. The inset shows the region of the 540\,keV peak in linear scale. There are wide enough regions on both sides of the peak where the background for the peak area determination could be fixed with good confidence. The small peak to the left of the 540\,keV peak is the 534.3\,keV peak from the decay of $^{156}$Tb. This isotope has been produced by the $^{153}$Eu($\alpha$,n)$^{156}$Tb reaction on the 0.8\,\% $^{153}$Eu impurity of the target material. As in the case of the spectrum shown, this peak was always very weak and could be well resolved from the 540.2\,keV one.

\section{Data analysis}
\label{sec:analysis}

The half-life of $^{154}$Tb$^{m1}$ has been determined by fitting the exponential decay curve to the measured and dead-time corrected 540.2\,keV peak area as a function of time. After the linearization of the exponential decay function, the analytic solution of the least square method could be used (see Ref. \cite{Leo94}). 

\begin{figure}
\centering
\resizebox{0.8\textwidth}{!}{\rotatebox{270}{\includegraphics{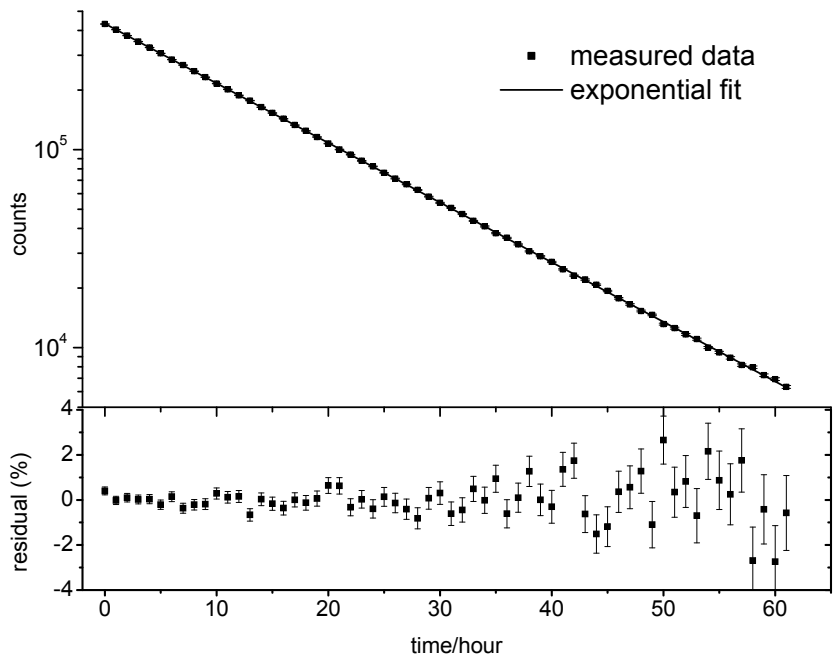}}}
\caption{\label{fig:decay} Decay curve of the source \#5}
\end{figure}

Figure \ref{fig:decay} shows the decay curve of source \#5 with the measured points and the fitted exponential function as well as the percentage residual in the lower panel. 

The results of the fits for the five sources are listed in Table \ref{tab:results}. The results obtained for $^{66}$Ga are also listed. This source was used to check one possible systematic uncertainty (see Sec. \ref{sec:uncertainties}). The quoted errors in the table are statistical only, systematic uncertainties will be discussed in Sec. \ref{sec:uncertainties}. From the listed reduced $\chi^2$ values one can conclude that the quality of the fit was good in all cases. In those two cases when the $\chi^2_{red}$ value was higher than one, the uncertainty of the half-life has been multiplied by $\sqrt{\chi^2_{red}}$. 

\begin{table}
\centering
\caption{\label{tab:results} Half-life results. The given uncertainties are statistical only. The results of $^{66}$Ga used for consistency check are also listed.}
\begin{tabular}{@{\extracolsep{5mm}}lcccc}
\hline
Source & \multicolumn{2}{c}{$^{154}$Tb$^{m1}$} & \multicolumn{2}{c}{$^{66}$Ga} \\
\cline{2-3}\cline{4-5}
  & T$_{1/2}$/h & $\chi^2_{red}$  & T$_{1/2}$/h & $\chi^2_{red}$ \\
\hline
\#1 & 9.993\,$\pm$\,0.106 & 0.88 & 9.367\,$\pm$\,0.098 & 0.97\\
\#2 & 10.008\,$\pm$\,0.033 & 0.85 & 9.496\,$\pm$\,0.168 & 1.21\\
\#3 & 9.984\,$\pm$\,0.019 & 0.98 & 9.520\,$\pm$\,0.202 & 0.81\\
\#4 & 10.008\,$\pm$\,0.029 & 0.90 & 9.400\,$\pm$\,0.176 & 0.91\\
\#5 & 9.994\,$\pm$\,0.006 & 1.31 & \multicolumn{2}{c}{yield too small}\\
\hline
weighted average &  9.994\,$\pm$\,0.006 & & 9.414\,$\pm$\,0.071 \\
literature value & 9.4\,$\pm$\,0.4 & & 9.49\,$\pm$\,0.07\\
\hline
\end{tabular}
\end{table}

\subsection{Systematic uncertainties}
\label{sec:uncertainties}

A number of possible systematic uncertainties may influence the obtained half-life results. First of all, any change of the detection efficiency in the course of the measurement leads to the alteration of the measured half-life. Such a change can be caused either by geometrical effect (e.g. changes in the source detector distance) or instabilities of the intrinsic efficiency of the detector. This possibility can be tested by comparing the results of several measurements. As can be seen in Table \ref{tab:results} the results of the five measurements are in perfect agreement, the standard deviation of the points is as low as 0.1\,\%. However, to give a conservative estimation, the difference between the highest and the lowest value, 0.24\,\%, has been adopted for the detection efficiency systematic uncertainty.

The reliable dead-time determination of the data acquisition system is also of high importance. The five sources have been measured under different conditions regarding the initial dead-time of the system and gave consistent results. Nevertheless, the dead-time provided by the multichannel analyzer has been tested using a pulse generator. The signals from the pulse generator have been fed into the counting system using the test input of the detector preamplifier. The observed number of pulser signals in the spectrum has been compared with the sent signals (measured by a counter) and from the ratio the dead-time could be calculated. It was found that the dead-time provided by the acquisition system was precise within about 0.3\,\% in the dead-time range encountered in the present experiment. This 0.3\,\% uncertainty translates into less than 0.1\,\% uncertainty of the half-life value, therefore 0.1\,\% has been adopted for the dead-time determination systematic uncertainty.

The effect of detection efficiency changes and dead-time determination can also be studied using a reference source with precisely known half-life. The half-life of the reference source should preferably be close to that of the investigated isotope. The backing of the Eu targets contained some copper impurity. During $\alpha$-bombardment the $^{63}$Cu($\alpha$,n) reaction takes place with high cross section producing $^{66}$Ga. The half-life of this isotope is 9.49\,h\,$\pm$\,0.07\,h \cite{NDS83}, similar to that of $^{154}$Tb$^{m1}$. The decay of this isotope could well be seen in all but one samples. The half-life of $^{66}$Ga was determined by measuring the 1039\,keV $\gamma$-radiation following its decay. The obtained results are also listed in Table \ref{tab:results}. The values obtained on the different sources are again in good agreement, and their weighted average is in agreement with the literature value. This independently supports the findings about the efficiency and dead-time related systematic uncertainties.

\subsection{Systematic uncertainties from the m2 and ground state decay of $^{154}$Tb}
\label{sec:Tb_uncertainties}

In the case of $^{154}$Tb a very important source of uncertainty in the m1 isomer half life is the possible contribution of the m2 isomer and the ground state leading to decays through the 540\,keV transition. The m2 isomer is known to decay by isomeric transition to the m1 state providing a continuos feeding of that state. The m2 decay branching ratio to the m1 state is, however, only 1.8\,\%\,$\pm$\,0.6\,\% \cite{Lau73}. On the other hand, the m2 isomer is only very weakly populated by the $^{151}$Eu($\alpha$,n)$^{154}$Tb reaction. Gavriljuk et al. \cite{Gav90} found that the cross section ratio to the m2 and m1 states is $\sigma_{m2}/\sigma_{m1}$\,=\,0.019\,$\pm$\,0.003 at 15.6\,MeV $\alpha$-energy. We have also calculated this ratio by measuring and comparing the characteristic $\gamma$-radiations for the two isomers. We found $\sigma_{m2}/\sigma_{m1}$ values in the range between 1.2\,\% and 2.7\,\% for the measured 5 different $\alpha$-energies. The combination of the low cross section ratio and the low isomeric transition decay ratio results in a maximum of 0.08\,\% feeding of the m1 state by the m2 decay (here a conservative upper limit of two sigma deviation from the isomeric transition literature value was taken). This feeding influences the determined m1 half-life by at most 0.1\,\% and therefore this value is taken as the systematic uncertainty from the m2 feeding.

Our half-life determination was based on the assumption that the 540.2\,keV $\gamma$-radiation comes exclusively from the m1 decay, as measured by e.g. Sousa et al. \cite{Sou75}. A small contribution of the ground state decay to this $\gamma$-line can, however, not be excluded. This is especially important because our analysis shows that the ground state is strongly populated not only by the decay of the m1 state but also directly by the ($\alpha$,n) reaction (typically $\sigma_{gs}/\sigma_{m1}$\,$\simeq$\,0.3). 

\begin{figure}
\centering
\resizebox{0.8\textwidth}{!}{\rotatebox{270}{\includegraphics{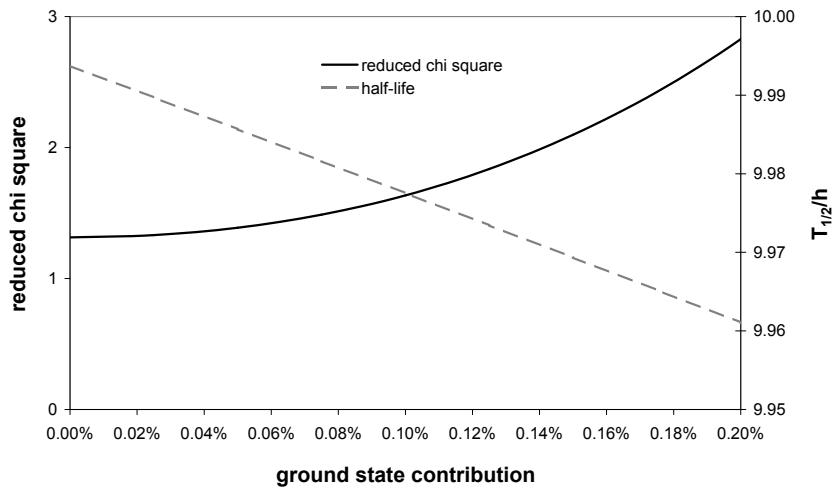}}}
\caption{\label{fig:chisquare} Results of the chi square analysis on the ground state decay contribution.}
\end{figure}

To test this possibility, a chi square analysis has been performed. A hypothetical ground state decay contribution to the 540.2\,keV $\gamma$-line has been assumed and the goodness-of-fit was checked by the reduced chi square value as a function of the ground state decay contribution. The results of this test can be seen in Fig. \ref{fig:chisquare}. The minimum of the chi square curve is at zero ground state contribution to the 540.2\,keV $\gamma$-line. This confirms the result of \cite{Sou75} that this line can be associated solely with the m1 isomer. At the point where the reduced chi square increases by one, the obtained half life changes by 0.27\,\%. This value is taken for the systematic uncertainty from ground state decay contribution.

\section{Results and conclusion}
\label{sec:results}

The adopted value for the half-life of the $^{154}$Tb$^{m1}$ isomer and its statistical uncertainty is calculated as the weighted average of the five measured samples. This gives 9.994\,h\,$\pm$\,0.006\,h. The total uncertainty is the quadratic sum of the statistical uncertainty and the following partial systematic uncertainties: detection efficiency (0.24\,\%), dead-time determination (0.1\,\%), m2 isomer feeding (0.1\,\%) and ground state decay contribution (0.27\,\%). The final result is T$_{1/2}$\,=\,9.994\,h\,$\pm$\,0.039\,h with a total uncertainty of 0.4\,\%.

The half-life value obtained in the present work is significantly higher than the recommended value in the literature (9.4\,h\,$\pm$\,0.4\,h \cite{NDS85}) and its uncertainty is one order of magnitude lower. It should be pointed out that our value is in agreement with the results of \cite{Vyl72}, \cite{Ber83} and \cite{Ned76}, in marginal agreement with \cite{Sou75} but in serious disagreement with \cite{Lau73}. Since our value is more precise than any of the previous results and it is in agreement with the available most precise values, the result of our measurement is recommended as the adopted value and the quandary of the compilation cited in Sec. \ref{sec:introduction} seems to be solved. With the high precision half-life of $^{154}$Tb$^{m1}$ determined in the present work, a systematic uncertainty in the $^{151}$Eu($\alpha$,n)$^{154}$Tb cross section measurement is significantly reduced.

\section*{Acknowledgments}

This work was supported by ERC Grant 203175,
GVOP-3.2.1.-2004-04-0402/3.0. and OTKA (K68801,
T49245). Gy. Gy. is a Bolyai fellow. G. R. is supported by the Visegr\'ad fund.

\end{document}